\documentclass[11pt,twoside]{article}
\usepackage{ldp5,flushrt,epsfig}

\newcommand{\etal}{et al.\ }
\newcommand{\eg}{e.g.,\ }
\newcommand{\ie}{i.e.,\ }

\begin{document}

\setcounter{page}{1}

\title{AUTOMATED STELLAR SPECTRAL CLASSIFICATION AND PARAMETERIZATION 
FOR THE MASSES}

\author{Ted von Hippel, Carlos Allende Prieto, \& Chris Sneden}

\affil{Department of Astronomy, University of Texas, Austin, TX 78712}

\GARfoot{-10}{Ted von Hippel}
{Dept. of Astronomy}
{University of Texas}{Austin, TX 78712 }{ted@astro.as.utexas.edu}
\vspace{.3in}

\markboth{\hspace{0.5in}{\rm Ted von Hippel, Carlos Allende Prieto, \& Chris Sneden}
\hspace{\fill}{\rm }}{{Automated Stellar Spectral Classification and 
Parameterization for the Masses}}

\abstract{
Stellar spectroscopic classification has been successfully automated by a
number of groups.  Automated classification and parameterization work best
when applied to a homogeneous data set, and thus these techniques
primarily have been developed for and applied to large surveys.  While
most ongoing large spectroscopic surveys target extragalactic objects,
many stellar spectra have been and will be obtained.  We briefly summarize
past work on automated classification and parameterization, with emphasis
on the work done in our group.  Accurate automated classification in the
spectral type domain and parameterization in the temperature domain have
been relatively easy.  Automated parameterization in the metallicity
domain, formally outside the MK system, has also been effective.  Due to
the subtle effects on the spectrum, automated classification in the
luminosity domain has been somewhat more difficult, but still successful.
In order to extend the use of automated techniques beyond a few surveys,
we present our current efforts at building a web-based automated stellar
spectroscopic classification and parameterization machine.  Our proposed
machinery would provide users with MK classifications as well as the
astrophysical parameters of effective temperature, surface gravity, mean
abundance, abundance anomalies, and microturbulence.}

\section{{A Brief History of Automated Classification}}
\nopagebreak

Current or planned large-scale surveys, such as the Sloan Digital Sky
Survey (York \etal 2000) or the GAIA mission (scheduled for launch around
2011), have led to increased interest in automated spectral classifiers
(\eg Bailer-Jones 2000).  There are many other reasons to develop
automated classifiers, not the least of which are the homogeneity of the
results and the repeatability of the process.  Automated spectral
classification of stars goes back decades.  For instance, in an early
attempt Jones (1966) fit a few major stellar lines and correlated these
indexes with MK type (Morgan, Keenan, \& Kellman 1943).  Malyuto \&
Shvelidze (1994) later developed this technique further.  Unfortunately,
the line fitting technique suffers from the disadvantage that one has
first to know the approximate stellar type before determining which lines
to fit, otherwise very different features will be found at the same
wavelengths.  Kurtz (1983; see also LaSala 1994) developed a minimum
vector distance technique that matched spectra to a library of standards
weighting the comparison to different spectral regions for different types
of stars.  The minimum vector distance technique has had some success --
classifying stellar spectral types to within $\sigma$=2.2 spectral
subtypes -- but refining this technique is cumbersome since the weighting
vectors need to be carefully established, yet they vary as a function of
spectral type and luminosity class.

In the middle of the last decade four independent groups (Gulati \etal
1994; von Hippel \etal 1994; Vieira \& Ponz 1995; Weaver \& Torres-Dodgen
1995) began to successfully apply artificial neural networks (ANNs) to the
spectral classification problem.  Neural networks are trained to yield
classifications that are identical to those previously assigned, and thus
have the advantage that the ANN builder need not become a classification
expert, but rather can rely on the true experts in the field, via their
many previous classifications.  For example, von Hippel \etal built an ANN
using the objective prism spectra and classifications of Houk (1982, and
references therein), who had at that time already classified more than
10$^5$ stars.  This first generation of neural networks were applied to
low resolution ultraviolet (Vieira \& Ponz 1995) or optical (the other
three studies) spectra and achieved $\sigma$=0.6 spectral subtypes and
$\sigma$=0.35 luminosity classes for A stars and returned E($B-V$) (Weaver
\& Torres-Dodgen 1995), or $\sigma < 2$ subtypes over the broad range of
O3 to M4 stars.  These studies validated the ANN approach and indicated
its tremendous potential.

The second generation of ANN spectral classification studies (Weaver \&
Torres-Dodgen 1997; Bailer-Jones, Irwin, \& von Hippel 1998; Singh,
Gulati, \& Gupta 1998; Weaver 2000) focused on increasing sample size and
moving on to two-dimensional classification.  At this point spectral type
classification became very good, with $\sigma$=0.5--0.7 subtypes.
Luminosity classification quality varied from $\sigma \approx 0.3$
luminosity classes (Weaver \& Torres-Dodgen 1997) to a statistically
reliable luminosity classification for dwarfs and giants, though not
sub-giants (Bailer-Jones \etal 1998).  Interestingly, Weaver (2000) also
showed that he could provide two dimensional classification for both
components of artificial binaries!

A few ANN studies (Bailer-Jones \etal 1997; Snider \etal 2001) moved from
MK stellar classification to parameterization of astrophysical parameters
($T_{\rm eff}$, log($g$), [Fe/H]).  Although the philosophy of
classification and parameterization are different, they strive to serve
the same community.  Classification seeks to place any program star within
the framework defined by a series of standards.  Since the MK
classification system is so widely used and the connection to stellar
parameters such as absolute magnitude, surface temperature, and mass, are
in general well known, this has been a productive route for studying
individual stars, local stellar populations, and Galactic structure.
Stellar parameterization seeks to skip the initial step of MK
classification and directly determine atmospheric parameters.  The cost is
that the results are model dependent, but in many parts of the HR diagram
model spectra look very much, though not exactly, like real stellar
spectra.  In addition, model atmospheres can be easily constructed for
subsolar metallicity, and therefore these can be applied to spectra which
would not be possible to classify on the MK system.  Bailer-Jones \etal
(1997) passed their objective prism spectra through ANNs trained on
stellar atmosphere models and derived a detailed mapping between MK
classifications and the Kurucz (1979, 1992) model atmosphere set they used
(as implemented with the program SPECTRUM by Gray \& Corbally 1994).  They
also reported that the mean metallicity in the solar neighborhood, as
represented by Houk's (1982) objective prism spectra, is slightly
subsolar, at [Fe/H] = $-0.2$. Snider \etal (2001) turned the problem
around and trained ANNs on real stellar spectra using atmospheric
parameters previously derived from fine abundance analysis work in the
literature, now in three-dimensional parameter space.  They achieved
$\sigma$($T_{\rm eff}) \approx 150$ K, $\sigma$(log($g$)) $\approx$ 0.33
dex, and $\sigma$(Fe/H) $\approx$ 0.2 dex.

\section{{A Few Lessons Learned}}
\nopagebreak

Here we offer a few comments on lessons we have learned in applying ANNs
to the problems of spectral classification and parameterization:

\begin{itemize}

\item Classification in spectral type and parameterization in $T_{\rm
eff}$ are easy, the ANNs have little trouble finding a good global
solution, and the results are generally precise to less than a spectral
subtype or 200 K.

\item Luminosity classification and log($g$) parameterization are
possible, but more difficult.  Both require spectra with an adequate
combination of resolution, wavelength coverage, and S/N.  This spectral
quality is just achieved with classical MK objective prism resolution and
wavelength coverage.

\item ANNs are best suited to stellar classification/parameterization when
a single wavelength range and resolution are used for a particular ANN.

\item If low S/N spectra are used, besides the higher random errors, ANNs
may make systematic errors unless they have been trained on low S/N
spectra (Snider \etal 2001).

\item Supervised ANNs treat spectra as patterns with a known correlation
between those patterns and answers, and attempt to learn that
relationship.  The better one can homogenize the training data so that the
ANN does not find spurious correlations between the input catalog and the
answers the better results one achieves.  Larger catalogs always help in
this regard, as do uniform data sets.  Spurious correlations caused by
real astrophysics can also be a problem.  As an example of this
phenomenon, the Houk catalog is magnitude-limited with $V_{\rm limit}$ =
10--11.  This magnitude limit creates a correlation between spectral type
and luminosity class, \ie the catalog contains mostly early type dwarfs
and late type giants.  Without proper scrutiny one might believe a trained
ANN has achieved true luminosity classification using such a catalog when
the ANN could have learned to classify luminosity statistically based on
spectral type.

\item Principle Component Analysis (PCA) can be used to compress stellar
spectra or to remove spurious signals (\eg Storrie-Lombardi \etal 1995).
PCA also bears a strong resemblance to ANNs (Lahav 1995), and is a good
pedagogical tool to gaining a heuristic understanding of ANNs.  By
creating a series of vectors, a linear combination of which will recreate
any star in the library, PCA recasts a stellar spectral library in much
the same way as the hidden nodes in a single hidden layer ANN.

\item Spectral classification errors are a function of spectral type,
based largely on the number of examples and variance within a given type
or range of types.  For example, Bailer-Jones \etal (1998) found
$\sigma$(SpT)=0.5 for B3 to A0 stars and $\sigma$(SpT)=0.8 for F3 to G1
stars.  The lower errors for the B and A stars are probably the result of
reduced sensitivity of their spectra to abundance differences.

\item It is easiest to build multiple ANNs, each specializing in a
specific dimension, when solving multi-dimensional spectral classification
or parameterization problems.  Not only are the ANNs more likely to
converge on a good global solution, but less data are required for this
approach.  For example, Snider \etal (2001) built ANNs specializing in
each of $T_{\rm eff}$, log($g$), and [Fe/H].  Weaver (2000) has also found
it helpful to use one ANN for initial rough classifications, followed by
specialist ANNs, trained on a limited spectral type ranges, to refine the
classifications.

\end{itemize}

\section{{Proposed Classification and Parameterization for the Masses}}
\nopagebreak

Can we build automated spectral classifiers for general use?  For some
time Bob Garrison has pointed out that we could build stellar classifiers
for particular spectrographs.  Users of such spectrographs might have a
near real-time reduction pipeline, immediately following which they would
receive a stellar classification.  This is certainly possible.  In fact,
if the data were obtained in a standard manner, the only reduction steps
required prior to ANN classification would be spectral extraction and
wavelength calibration.

In practice, neither we nor, to the best of our knowledge, any other group
has taken this approach.  The difficulties are not technical, but rather
the time-consuming nature of building multiple such classification
machines for the many possible spectrographs in use by stellar
spectroscopists.  Certainly, if a stellar spectroscopic survey of
sufficient size were to be undertaken which would create a uniform data
set of sufficient quality, we and others would be motivated to build a
tailor-made stellar classification or parameterization machine for that
instrument/survey.

We propose instead a thematically related approach.  Instead of building
ANN classification/parameterization machines for particular spectrographs,
we propose to build such machines for particular combinations of
wavelength coverage, resolution, and S/N, and make these available for use
via the web.  It would be up to the user to process their data onto a
linear flux and wavelength scale at one of the resolutions and wavelength
ranges supported by our web site algorithms.  Users would upload their
spectra, run the classification or parameterization ANNs, and receive a
spectral type and luminosity class and/or the stellar astrophysical
parameters, along with the associated uncertainties.

We hope to develop such a tool first by beginning with stellar
parameterization based on model atmospheres.  Our entire approach would be
modular and would initially support a single resolution and wavelength
range, while covering the parameter range 4500 $\leq T_{\rm eff} \leq
8000$, 2 $\leq$ log($g$) $\leq$ 5, $-4.5 \leq$ [Fe/H] $\leq +0.5$, and 0
$\leq$ microturbulence $\leq$ 1/2 the stellar rotation break-up velocity.
Our initial resolution and wavelength range have not been finalized, but
would probably be $R \approx$ 2000 and 150 \AA\ around H$\beta$,
respectively.  This would allow anyone with higher resolution spectra or
spectra with a broader range of wavelengths to take advantage of our first
automated parameterization algorithms.  As we move to support a wider
range of spectral resolutions and wavelength coverages we also intend to
add modules to increase our effective temperature range to $T_{\rm eff}$ =
50,000, decrease our surface gravity range to log($g$) = 1, include
different relative O, Mg, and Ca abundances, and begin spectral
classification on the MK system for stars of near solar metallicity.
Eventually we hope to push the stellar parameters into the M, L, and T
dwarf regimes.  We recognize that model atmospheres are never perfectly
accurate and that they improve with time.  From time to time, where
meaningful advances have been made for a particular range of stellar
atmospheres, we will upgrade our parameterization modules.

Our anticipated users are spectroscopists doing fine-abundance analysis
who want a starting point or a sanity check on their result, those
conducting surveys who need classifications or parameterizations for
statistical or pre-selection purposes, and those wanting independent
determinations of the classifications/parameters for their program stars
for a wide variety of studies.

\vfill\eject

\acknowledgments

We wishes to thank Bob Garrison, Richard Gray, and Coryn Bailer-Jones for
thought-provoking discussions.  We gratefully acknowledge major financial
support for this work from the State of Texas Advanced Research Program.


\begin{references}

Bailer-Jones, C. A. L. 2000 A\&A 357, 197

Bailer-Jones, C. A. L., Irwin, M., Gilmore, G. \& von Hippel, T. 1997
MNRAS 292, 157

Bailer-Jones, C. A. L., Irwin, M. \& von Hippel, T. 1998 MNRAS 298, 361

Gray, R. O. \& Corbally, C. J. 1994 AJ, 107, 742

Gulati, R. K., Gupta, R., Gothoskar, P. \& Khobragade, S. 1994 ApJ, 426,
340

Houk, N. 1982 University of Michigan Catalogue of Two-Dimensional Spectral
Types for the HD Stars, Vol. 3

Jones, D. H. P. 1966 R. Obs. Bull. 126, 219

Kibblewhite, E. J., Bridgeland, M. T., Bunclark, P. S. \& Irwin, M. J.,
1984 (ed.) Klinglesmith D.A. NASA-2317, Astronomical Microdensitometry
Conference.  NASA, Washington D.C., p. 277

Kurtz, M. J. 1983 (ed.) Garrison,  B. F., The MK Process and Stellar 
Classification, David Dunlop Observatory, Toronto, p. 136

Kurucz, R. L. 1979 ApJS, 40, 1

Kurucz, R. L. 1992 (ed.) B. Barbuy \& A. Renzini, Stellar Populations
of Galaxies, IAU Symp. 149, Dordrecht, Kluwer, 225

Lahav, O. 1995 Vistas in Astronomy 38, 251

LaSala, J. 1994 (ed.) C. J. Corbally, R. O. Gray, \& R.  F. Garrison, The
MK Process at 50 Years: A Powerful Tool for Astrophysical Insight, ASP
Conf. Ser. 60, San Francisco, ASP, 312

Malyuto, V. \& Shvelidze, T. 1994 (ed.) C. J. Corbally, R. O. Gray, \& R.
F. Garrison, The MK Process at 50 Years: A Powerful Tool for Astrophysical
Insight, ASP Conf. Ser. 60, San Francisco, ASP, 344

Morgan, W. W., Keenan, P. C. \& Kellman, E., 1943, An Atlas of Stellar
Spectra, University of Chicago Press, Chicago

Singh, H. P., Gulati, R. K. \& Gupta, R. 1998 MNRAS 295, 312

Snider, S., Allende Prieto, C., von Hippel, T., Beers, T. C., Sneden, C.,
Qu, Y. \& Rossi, S. 2001 ApJ 562, 528

Storrie-Lombardi, M. C., Irwin, M. J., von Hippel, T. \& Storrie-Lombardi,
L. J. 1995 Vistas in Astronomy 38, 331

Vieira, E. F. \& Ponz, J. D. 1995 A\&AS 111, 393

von Hippel, T., Storrie-Lombardi, L. J., Storrie-Lombardi, M. \& Irwin,
M. J. 1994 MNRAS 269, 97

Weaver, W. B. 2000 ApJ 541, 298

Weaver, W. B. \& Torres-Dodgen, A. V. 1995 ApJ 446, 300

Weaver, W. B. \& Torres-Dodgen, A. V. 1997 ApJ 487, 847

York, D. G. \etal 2000 AJ 120, 1579

\end{references}
\end{document}